\documentclass[useAMS,usenatbib]{mn2e}
\usepackage{graphicx}
\usepackage{float}
\usepackage{color}
\usepackage{tabularx}
\def\kms{km\,s$^{-1}$}

\def\Ha{H$\alpha$}
\def\Hb{H$\beta$}

\def\CaII{Ca\,{\sc ii}}

\def\OI{w}
\def\HI{H\,{\sc i}}

\def\SiII{Si\,{\sc ii}}

\def\HeI{He\,{\sc i}}
\def\FeII{Fe\,{\sc ii}}

\def\NII{N\,{\sc ii}}

\def\TiII{Ti\,{\sc ii}}

\def\LL{$\lambda\lambda$}
\def\kms{km\,s$^{-1}$}

\def\Ha{H$\alpha$}
\def\Hb{H$\beta$}

\def\CaII{Ca\,{\sc ii}}

\def\OI{O\,{\sc i}}

\def\SiII{Si\,{\sc ii}}
\def\TiII{Ti\,{\sc ii}}

\def\HeI{He\,{\sc i}}
\def\FeII{Fe\,{\sc ii}}

\def\NII{N\,{\sc ii}}

\def\msun{M$_{\odot}$}
\onecolumn
\title[The first month of evolution of the type II-P SN~2013ej]
{The first month of evolution of the slow rising type II-P SN~2013ej in M74\thanks{This paper is based on observations obtained with the following telescopes: LCOGT Network of 1m and 2m Telescopes,  NTT(184.D-1140,188.D-3003), IRTF}}

\author[Valenti et al.]{S. Valenti$^{1,2}$
\thanks{e--mail: svalenti@lcogt.net},
D. Sand$^{3}$, A. Pastorello$^{4}$,  M. L. Graham$^{1,2}$, D. A. Howell$^{1,2}$, 
\and
 J. Parrent$^{1,5}$,  L. Tomasella$^{4}$, P. Ochner$^{4}$, M. Fraser$^{6}$, S. Benetti$^{4}$,  F. Yuan$^{7}$,     
\and
S. J. Smartt$^{6}$,  J. R. Maund$^{6}$, I. Arcavi$^{8}$, A. Gal-Yam$^{8}$, C. Inserra$^{6}$, D. Young$^{6}$ \\
$^{1 ~}$ Las Cumbres Observatory Global Telescope Network, 6740 Cortona Dr., Suite 102, Goleta, CA 93117, USA \\
$^{2 ~}$ Department of Physics, University of California, Santa Barbara, Broida Hall, Mail Code 9530, Santa Barbara, CA 93106-9530, USA) \\
$^{3 ~}$ Physics Department, Texas Tech University, Lubbock, TX 79409, USA \\
$^{4 ~}$ INAF Osservatorio Astronomico di Padova, Vicolo dell'Osservatorio 5, 35122 Padova, Italy \\
$^{5 ~}$ Department of Physics and Astronomy, Dartmouth College, 6127 Wilder Laboratory, Hanover, NH 03755, USA \\
$^{6 ~}$ Astrophysics Research Centre,  School of Mathematics and Physics, Queens University Belfast, Belfast BT7 1NN, UK \\
$^{7 ~}$ Research School of Astronomy and Astrophysics, The Australian National University, Weston Creek, ACT 2611, Australia \\
$^{8}$ Department of Particle Physics and Astrophysics, The Weizmann Institute of Science, Rehovot 76100, Israel \\
}
\begin{document}

\date{Accepted .....; Received ....; in original form ....}
\maketitle

\begin{abstract}
We present early photometric and spectroscopic observations of
SN~2013ej, a bright type IIP supernova in M74. SN~2013ej is one of the
closest SNe ever discovered. The available archive images and the
early discovery help to constrain the nature of its progenitor. The
earliest detection of this explosion was on 2013 July 24.14 UT and our
spectroscopic monitoring began on July 27.73 UT, continuing almost
daily for two weeks with the LCOGT FLOYDS spectrographs.  Daily
optical photometric monitoring was achieved with the LCOGT 1m network,
and were complemented by UV data from \emph{SWIFT} and near-infrared
spectra from PESSTO and IRTF.  The data from our monitoring campaign
show that SN~2013ej experienced a 10-day rise before entering into a
well defined plateau phase. This unusually long rise time for a type
IIP has been seen previously in SN~2006bp and SN~2009bw.  A relatively
rare strong absorption blue-ward of \Ha{} is present since our
earliest spectrum. We identify this feature as \SiII{}, rather than
high velocity \Ha{} as sometimes reported in the literature.
\end{abstract}


\section{Introduction}
\label{parintroduction}

Type IIP supernovae (SNe) have been widely studied in the past. They
arise from progenitors that have retained their hydrogen and helium
layers before exploding as core-collapse (CC) SNe. Several progenitors
of type IIP SNe have been detected in archive images \citep[see ][ for
  a review]{2009ARA&A..47...63S}. These SNe are also used as standard
candles for cosmological studies \citep{Hamuy2001}. For this reason it
is important to investigate their diversity when new close-by type II
SNe are discovered.  We have this opportunity with SN~2013ej, a young
SN recently discovered in M74 (NGC 628).

SN~2013ej was discovered by the Lick Observatory Supernova Search
(LOSS) on 2013 July 25.45 UT, with the 0.76-m Katzman Automatic
Imaging Telescope (KAIT). The coordinates of the SN are $\alpha$ =
01h36m48.16s, $\delta$ = +15$^{o}$45'm31.0" \citep{Kim2013}.
Pre-discovery detections on July 25.38 and on July 24.125 were also
reported by Dhungana at al (2013) with the 0.45-m ROTSE-IIIb telescope
at McDonald Observatory and by C. Feliciano on the {\it Bright
  Supernovae}\footnote{http://www.rochesterastronomy.org/snimages/}
web site.  The SN was caught extremely young, as a non-detection was
reported by the All Sky Automated Survey for SuperNovae
\citep[ASAS-SN, ][]{atel5237} on July 23.54 UT\footnote{We will use
  this epoch (JD=2456497.0$\pm$1.0) as the reference date for the
  shock breakout}, less than one day prior to the first detection,
with a limit (though not very stringent) of $V$ $>$16.7 mag.  We
immediately triggered the robotic FLOYDS spectrograph mounted on the
Faulkes Telescope South (FTS) at Siding Spring Observatory and were
able to classify the transient on Jul. 27.70 UT as a young type II SN
\citep{Valenti2013}.  Located 92".5 E, 135" S from the core of
M74\footnote{A distance modulus of 29.79 mag (Fraser et al, in prep.),
  or 9.1 Mpc, is used in this paper.}, SN~2013ej is a very nearby
SN. We note that M74 hosted two other CC SNe, viz. SN~2002ap (SN Ic)
and SN~2003gd (SN IIP).

Deep, high resolution images of the host galaxy have been taken prior
to the explosion of SN~2013ej using the Hubble Space Telescope (HST)
and Gemini telescope. An analysis of these together with the
identification and characterization of a progenitor star is presented
in a companion paper (Fraser et al. 2013).  A bright source in both
the HST and Gemini images is identified by Fraser et al. (2013) as
coincident with the position of SN~2013ej. The object is detected from
the $U$ to the $i$ bands and does not have the colours of a single
stellar source. Fraser et al. (2013) show that it is likely a blend of
two, physically unrelated, stars and that the progenitor was
consistent with a red supergiant.

For the reasons mentioned above, SN~2013ej was promoted as a high priority 
target for LCOGT's ongoing, intensive effort to obtain high cadence optical 
photometry and spectroscopy of transient objects.  A detailed overview of 
the LCOGT facilities can be found in \cite{Brown2013}.  Along similar lines, 
SN~2013ej was also monitored by the PESSTO\footnote{Public ESO Spectroscopic 
Survey of Transient Objects, www.pessto.org/} collaboration.

\section{Observations}
Spectroscopic followup of SN~2013ej was primarily obtained using the twin 
FLOYDS spectrographs on FTS and the Faulkes Telescope North (FTN), at 
Haleakala, with a $\sim$1 day cadence.  FLOYDS has a prism cross-disperser 
which images first and second orders onto the chip, resulting in a very 
broad wavelength coverage of $\sim$3200-10000\AA.   Optical spectra were 
also obtained at the Asiago 1.22m telescope with the B\&C spectrograph.  
All optical spectra were reduced using standard IRAF routines.
An higher resolution spectrum was obtained on August 23rd with the 1.82m 
Copernico Telescope (Mt. Ekar, Asiago) equipped with an echelle spectrograph.
Two near-infrared (NIR) spectra were also collected.  A very early NIR 
spectrum (taken earlier than our first optical spectrum) was obtained 
with SpeX \citep{Rayner2003} on the NASA Infrared Telescope Facility (IRTF).
The SpeX data were reduced with the custom Spextool package \citep{Cushing2004}, 
and were corrected for telluric absorption with the software and prescription 
of \citet{Vacca2003}.  SOFI data were obtained during a PESSTO night and 
reduced with the PESSTO pipeline (Smartt et al in prep.)
(data available online (Tab. 1 and Tab. 2) 

Optical ($UBVRIgriz$) photometric follow-up of SN~2013ej was obtained 
with the nine 1-m telescopes of the LCOGT network, and in this paper 
we present data obtained up to one month after the SN discovery.  The 
data were reduced using a custom pipeline\footnote{The pipeline employs 
standard procedures ($pyraf$, $DAOPHOT$, $SWARP$) in a python framework. 
Point-spread function instrumental magnitudes were transformed to the 
standard Sloan Digital Sky Survey (SDSS) filter system (for ugriz) or 
\citet{1992AJ....104..340L} system (for UBVRI) via standard star observations 
taken during clear nights.} developed by the LCOGT SN team.
Additionally, {\it SWIFT} target of opportunity observations of SN~2013ej 
with the Ultraviolet/Optical Telescope \cite[UVOT;][]{Roming2005} were 
taken starting on 2013 July 30.87 UT.  {\it Swift} photometry, going 
from the $uw2$ filter ($\lambda_{c}$=1928\AA) through the $V$ filter 
($\lambda_{c}$=5468\AA), has been measured using an aperture of 3$"$, 
and following the approach of \cite{Brown2009a}.

\section{Analysis and Results}
\subsection{Spectroscopy}
\begin{figure*}
\begin{center}
\includegraphics[width=7.2in]{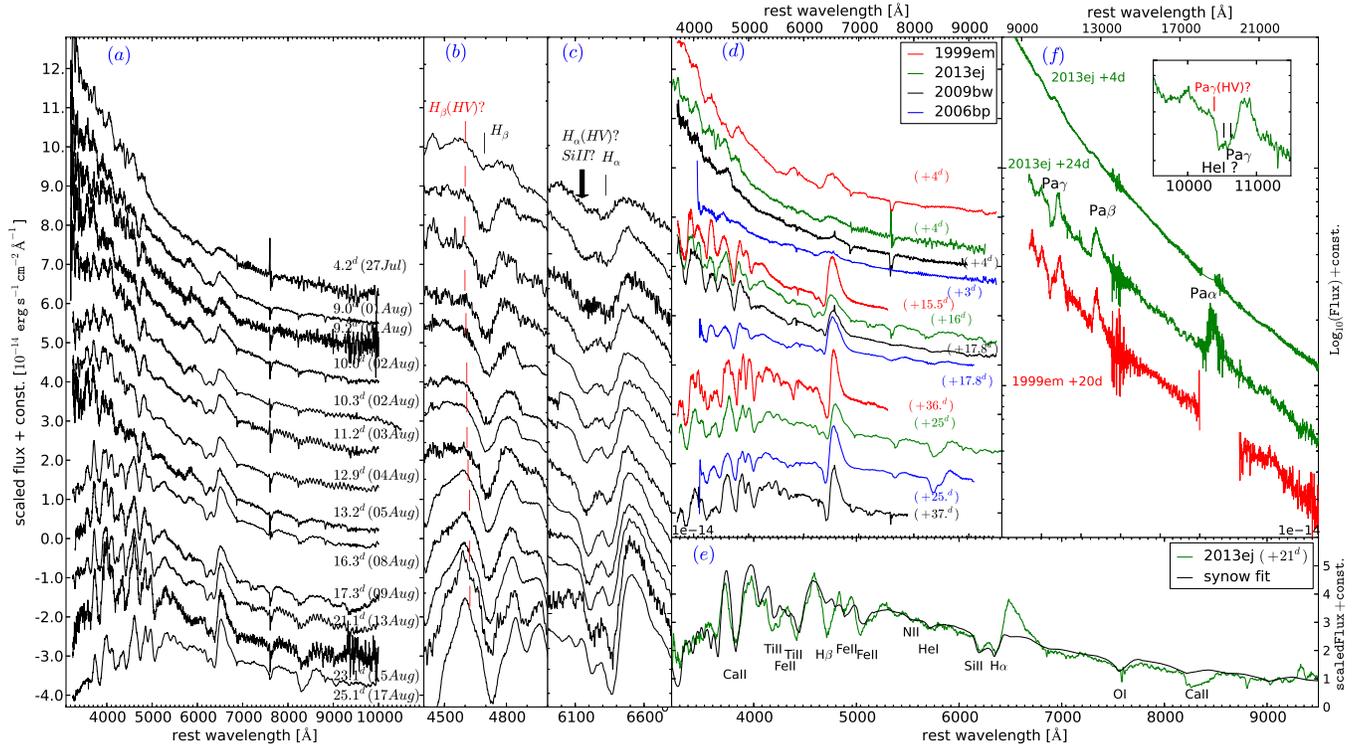}
\caption{\textbf{(a)}: Spectra series of SN~2013ej. \textbf{(b)}: \Hb{} evolution; 
\textbf{( c )} \Ha{} evolution; \textbf{(d)}: Comparison of spectra of SN~2013ej 
with those of other SNe II at three different epochs. \textbf{(e)}: 
Line identification: comparison between an observed spectrum of SN~2013ej 
(phase +20$^d$) and a spectral model obtained using $synow$ (see text). 
\textbf{(f)}: Infrared spectra of SN~2013ej and comparison with an early-phase 
spectrum of SN 1999em. The insert shows a detail of Pa-$\gamma$.}
\label{fig1}
\end{center}
\end{figure*}

The sequence of spectra collected for SN~2013ej is shown in Fig. \ref{fig1}a. 
In analogy with canonical type II SNe, at very early phases the spectra 
of SN~2013ej show a blue, almost featureless continuum. As the ejecta expand, 
the photospheric temperature declines, and a number of spectral 
lines appear, showing P-Cygni profiles that become more prominent with time. 
The earliest spectrum shows a blue continuum, with a black body temperature of 
15000 $K$.  Together  with weak and shallow Balmer-series lines, \NII{} and \HeI{} 
likely contribute to the absorption features at 5550 \AA{} and 5700\AA{}.
From very early phases, an absorption feature on the blue side of \Ha{} is detected 
(see Fig. \ref{fig1}c). This feature, observed in several young SNe II, has been 
widely discussed in the past, particularly for the classic type IIP SN~1999em. 
 The two identifications proposed to explain the dip are a high velocity (HV) 
component of \Ha{} (at $\sim$16,000 \kms{}) or, alternatively,  \SiII{} 6355\AA.
The evolution of this feature has been observed in several IIP SNe  (SN~2005cs,  
\citealt{2006MNRAS.370.1752P}; SN~2006bp,  \citealt{Quimby2007}; 
SN~2009bw \citealt{2012MNRAS.422.1122I} and SN~2012A, \citealt{Tomasella2013}).
It appears a few days after shock break-out, it reaches a maximum intensity 
at around three weeks after maximum light before eventually fading.
In Fig. \ref{fig1}d spectra at these three different epochs of SN~2013ej are 
compared with those of other SNe IIP  (a few days, 2 weeks and 3-4 weeks after 
the explosion respectively). In SN~1999em this absorption was visible since 
about one week post-explosion, and remained detectable for one additional week.
\cite{leonard2002} identified this absorption at early phases as \SiII{}, while 
\cite{Baron2000} favoured a HV hydrogen interpretation. 
\cite{2012MNRAS.422.1122I} also interpreted this feature as a high 
velocity hydrogen line in the type II SN~2009bw.\\
\cite{Chugai2007}, on the other hand, proposed a time dependent, dual interpretation.  
These authors attributed the shallow absorption in the \Ha{} blue wing visible at early 
phases in SNe type IIP to \SiII{}, but a similar blue-shifted notch seen at late times to 
HV hydrogen.  Indeed, the interaction with a typical red supergiant wind should result 
in the enhanced excitation of the outer layers of unshocked ejecta and the emergence 
of corresponding HV absorption \citep[see also][for a similar conclusion]{leonard2002}.
In the case of SN~2013ej, we note a similar evolution of this absorption feature to 
previous works, although a notch is visible in our first spectrum and becomes stronger 
with time. At day $+$17 the blue-shifted absorption is deeper than the regular \Ha{} 
absorption. We also note that a similar feature is not visible in the blue wing of \Hb{} 
(see Fig. \ref{fig1}b), as one would expect if it is due to HV hydrogen.

In order to confidently exclude the HV H interpretation for the 6100 \AA ~feature, 
we also present in Fig. \ref{fig1}f two NIR spectra obtained  on July 27th and  August 17th. 
The  first SpeX spectrum (+ 4$^d$) is one of the earliest NIR spectra ever published 
for a SN type IIP.  This shows a blue continuum, in addition in which only very 
shallow Paschen lines are detected. These become more prominent in the second spectrum, 
and show a broad absorption in Pa-$\gamma$ (see inset of Fig. \ref{fig1}f). 
While the red wing of the absorption is consistent with Pa-$\gamma$ at $\sim$ 8000 \kms{}
consistent with other hydrogen lines, the origin of the blue wing of this feature 
must wait for future work with time-series NIR spectroscopy. HV Pa-$\gamma$ at 
$\sim$ 15000 \kms{} (as suggested by the HV hydrogen identification in the optical) 
can be excluded.

A $synow$ \citep{2000PhDT.........6F} fit of the spectrum at 21 days is shown in Fig. 
\ref{fig1}e, and the comparison model was obtained with the following contributing  
ions: \FeII{},\TiII{},\HeI{},\HI{},\OI{}, \SiII{}, \CaII{},\NII{}. The \Hb{} feature is not well 
reproduced because of the non-LTE condition of the H-rich layers. Enhancing the 
optical depth of HI may help improving the match with the observed profile of \Hb{}, 
but this would make the \Ha{} absorption too prominent. Adding \SiII{} nicely 
reproduces the feature on the blue side of \Ha{} with no significant change in the 
$synow$ spectrum due to other \SiII{} lines. Also remarkable are the \FeII{} 
lines, first detected when the temperature goes below 9000 $K$, (as suggested by 
\cite{Dessart2006}). In particular, the \FeII{} line at 5166 \LL{}  appears two weeks 
after the core-collapse and is clearly detected one week later. 
In Fig. \ref{fig2}a we show the photospheric velocity where lines of different ions are 
forming. The velocity obtained for the \SiII{} identification is also consistent 
with those of other elements.

\subsection{Light curve}\label{sec:lc}
\begin{figure}
\begin{center}
\includegraphics[height=7.5in]{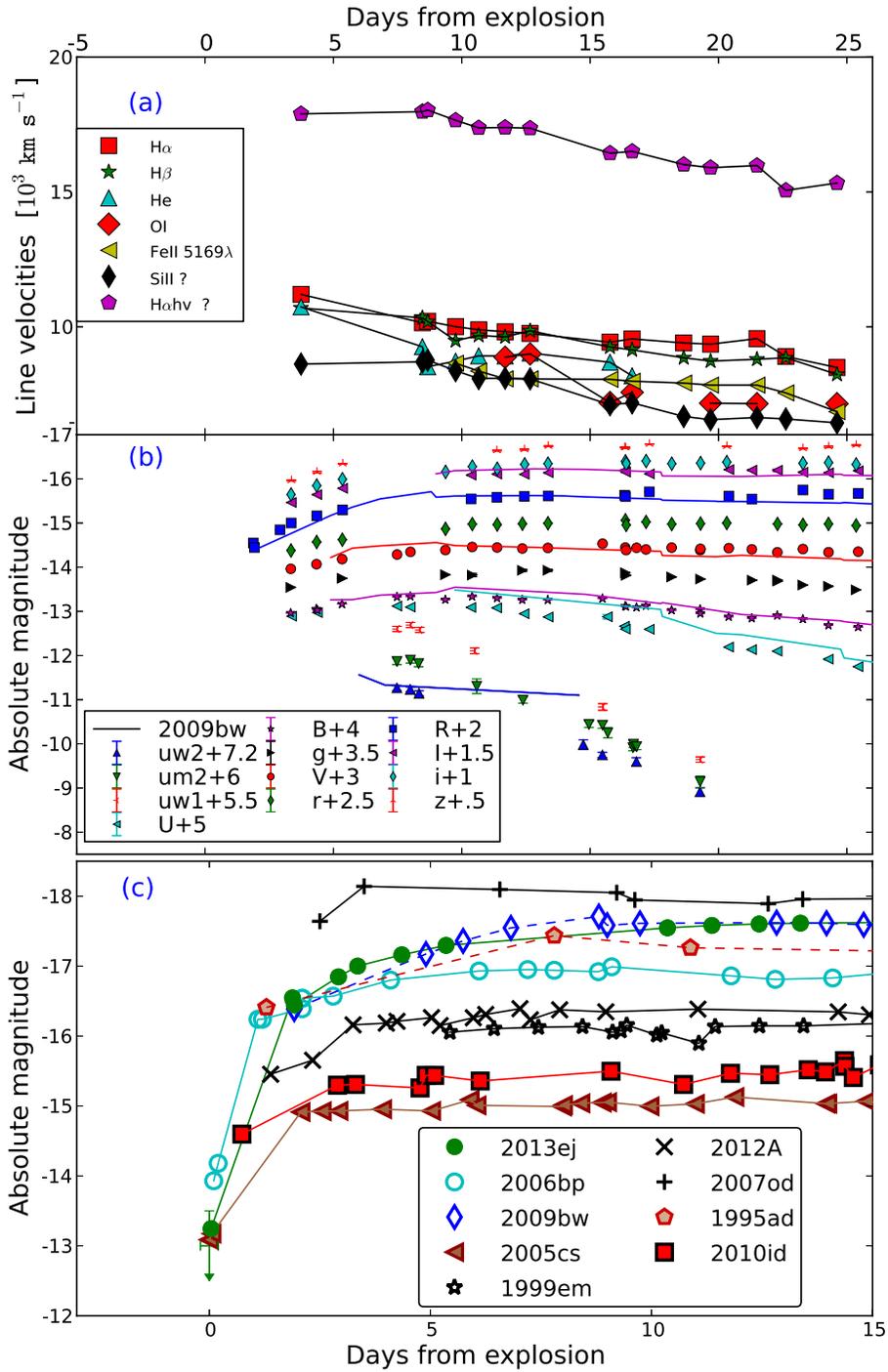}
\caption{\textbf{(a)}: Light curves of SN~2013ej in different bands (symbols), and 
comparison with the light curves of the twin SN~2009bw (solid lines).
\textbf{(b)}: Absolute R-band light curve of SN~2013ej compared with those of other 
type IIP SNe, including SN~2006bp, \citealt{Quimby2007}; SN~2009bw, 
\citealt{2012MNRAS.422.1122I}; SN~2005cs \citealt{2006MNRAS.370.1752P}; 
SN~2007od, \citealt{2011MNRAS.417..261I}; SN~2012A, 
\citealt{Tomasella2013}, SN~1995ad, \citealt{Inserra2013}.  
}
\label{fig2}
\end{center}
\end{figure}

The light curve of SN~2013ej in the $uw2$,$um2$,$uw1$,$UBVRIgriz$j filters is shown 
in Fig. \ref{fig2}b. SN~2013ej shows in all bands a relatively slow luminosity rise 
to the plateau. SNe II usually have a very fast rising lasting very few days.
Through the comparison of the early light curve of SN~2010id with that of SN~2006bp 
\citep{Quimby2007}, \cite{Gal-Yam2011} suggested a variety in the rising time of 
type II SNe, with some objects  showing a slow rise. Another SN II with a slow 
rise to peak is SN~2009bw \citep{2012MNRAS.422.1122I}. These three SNe are all quite 
bright ($M_{R}$ between -17 and -18 at peak (see Fig. \ref{fig2}c). In particular, 
SN~2009bw is almost identical to SN~2013ej in the light curve shape  and the 
absolute magnitude (see Fig. \ref{fig2}b). The rise time for type II SNe is 
usually connected to the radius of the progenitor at the time of its explosion. 
The slower the rise time, the more compact the progenitor. However the temperature 
evolution  of SN~2013ej (Sect. 3.1) seems to favour an extended progenitor. 
 
\subsection{Progenitor Radius}
In core-collapse SNe, soon after the shock break-out, the shock-heated envelope 
expands and cools down with different time scales depending on the initial 
progenitor radius, opacity and gas composition. Simple analytic functions have 
been developed  \citep{Waxman2007,Rabinak2011,Chevalier2011} to give a rough 
estimate of the radius of the progenitor of core-collapse SNe using the temperature 
evolution at early phases. For instance, the SN photosphere after the explosion of 
a red supergiant (RSG) remains at an higher temperature for a longer time than what 
occurs in a more compact blue supergiant (BSG). Making use of our spectro photometric 
data within one week from the shock-breakout and the eq. 13 of  \cite{Rabinak2011},  
we constrained the radius of the exploding star.  With an optical opacity of $k$=0.34 $cm^{2}~g^{-1}$ 
the temperature evolution of SN~2013ej is consistent with that of a progenitor with radius 400-600 $R_\odot$.

\begin{figure}
\begin{center}
\includegraphics[height=5.4in]{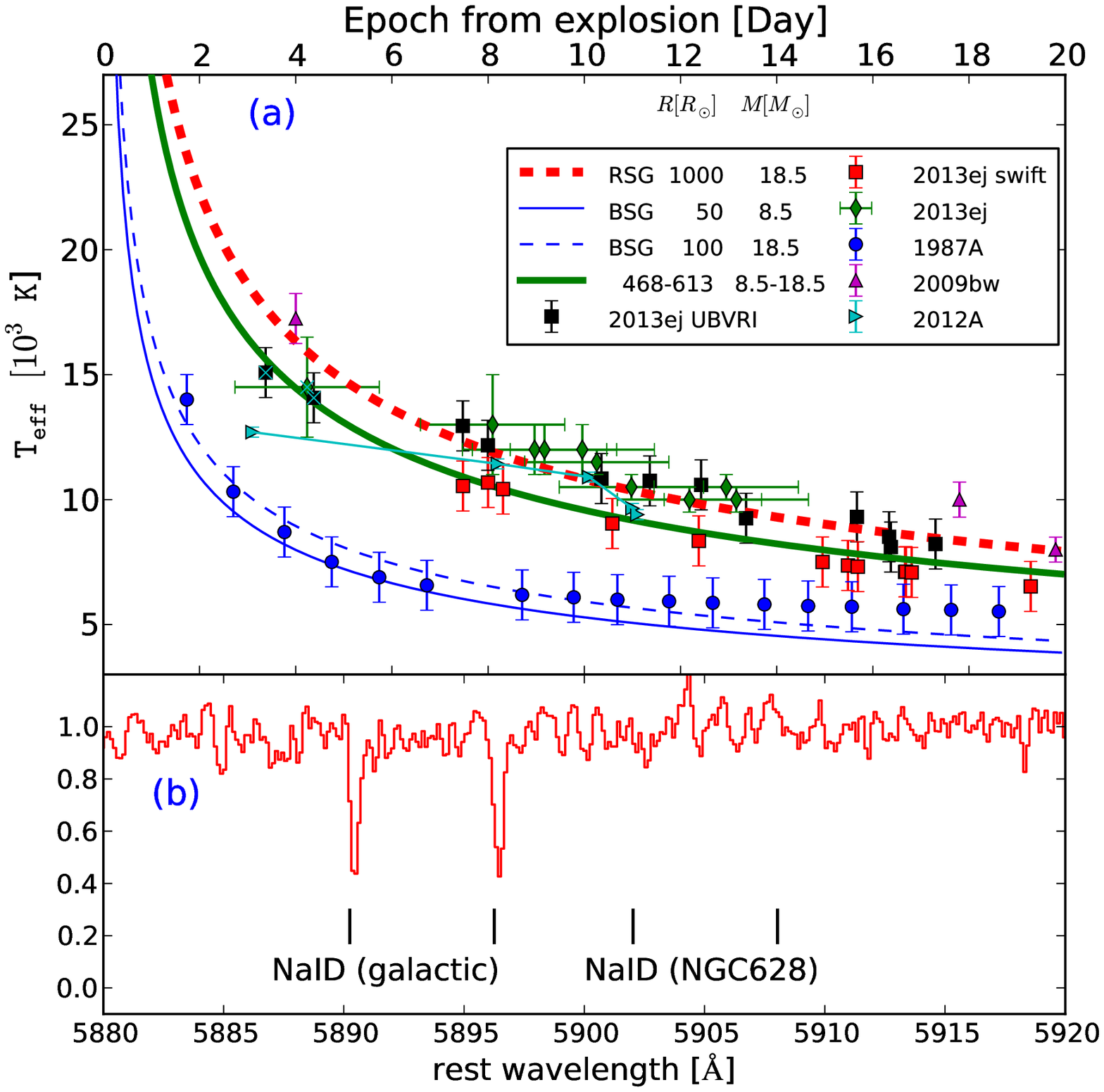}.
\caption{ \textbf{(a)}: Progenitor radius constraints using \protect\cite{Rabinak2011} 
for RSG (red line), BSG (blue line) and 2013ej best fit (green line). 
Comparison objects: SN~1987A  \citep{Menzies1987}, 
SN~2012A \citep{Tomasella2013}, SN~2009bw \citep{2012MNRAS.422.1122I};
\textbf{(b)}: Echelle spectrum of SN~2013ej obtained at the 1.8m asiago 
telescope on 24th August 2013. The Galactic NaID is clearly visible, while 
no NaID is detected at the redshift of M74. The x-axis is in the observed 
frame. \textbf{(c)}:  Photospheric velocity evolution for different elements 
in the ejecta of SN~2013ej.}
\label{fig3}
\end{center}
\end{figure}

 We remark that no host reddening was assumed in this calculation of the progenitor 
radius, which is consistent with the lack of NaID lines at the redshift of M74 
in the echelle spectrum obtained with the 1.82m Asiago telescope (see Fig. \ref{fig3}b). 
With the addition of some host reddening, the temperature and consequently the 
radius estimates would be larger. \\
The likely RSG progenitor star identified by Fraser et al. (2013) has a luminosity 
of $\log L/{\rm L_{\odot}} = 4.5-4.9$\,dex, which would imply radii of 400-800R$_{\odot}$ 
for M-type supergiant stellar effective temperatures. These radii are also consistent 
with that of the type IIP SN~2009bw \citep{2012MNRAS.422.1122I}, which shares several 
other characteristics with SN~2013ej, as we will discuss in the next section.\\
The inferred radius values should be confirmed with more detailed models, but here we 
may safely conclude that the temperature evolution of SN~2013ej is consistent with 
that expected in the explosion of an extended progenitor. In Fig. \ref{fig3}a we also 
present progenitor radius constraints for SN~2013ej utilizing our {\it SWIFT} and 
optical broadband photometry, which generally agree with our spectroscopic results.  
We note that the temperature evolution inferred using the {\it SWIFT} data is always 
lower than that derived from the optical data, suggesting that some UV line blanketing 
is already present at very early phases. We further compare the temperature evolution 
of SN~2013ej with those of other type II SNe. The photospheric temperatures and radii 
of SN~2013ej are similar to those of SN~2009bw and SN~2012A, while they are always 
higher than those of SN~1987A.

\section{Discussion and conclusion}
In this letter we present the early photometric and spectroscopic evolution 
of SN~2013ej, a luminous type II SN that exploded in the very nearby spiral 
galaxy M74. Data were obtained mostly with the 1m and 2m telescopes of the  
LCOGT network, complemented with data from the PESSTO collaboration. The most 
interesting characteristic of our sequence of early-time spectra of SN~2013ej is
a prominent absorption on the blue side of the \Ha{}. 
We interpret it  as due to \SiII{}, since there is no evidence of high 
velocity hydrogen features in proximity other Balmer or Paschen lines. 
This absorption dip is not particularly strong in other 
\emph{slow rising} type II SNe, although it was detected in SN~2009bw.
\cite{2012MNRAS.422.1122I} proposed it to be an evidence that SN~2009bw 
was interacting  with the H-rich CSM already at early phases.  
However, for the reasons mentioned above, we do not find an unequivocal evidence of 
interaction in  SN~2013ej - at least not before 30 days after explosion.
Nonetheless, we can not rule out that signatures of interaction 
between SN ejecta and CSM may appear at later stages, since this was observed in a 
number of type II SNe \citep{Chugai2007}. SN~2013ej shows an unusually slow rise 
to the light curve plateau. This property was observed  only in the type IIP SN~2006bp 
and SN~2009bw. All of them are more luminous than normal SNe II \citep{Patat1994}, 
but the statistics accumulated so far are poor.
Through modeling \citep{Zampieri2003,Pumo2011}, \cite{2012MNRAS.422.1122I}
estimated a progenitor mass of 11-15 \msun{} for SN~2009bw. If the early 
spectroscopic and photometric similarities of SN~2013ej with SN~2009bw, are 
confirmed at late phases, this mass range should also be considered plausible 
in the case of SN~2013ej, and  consistent with the progenitor mass inferred 
from the analysis of Fraser et al. (2013) (8-15.5 \msun{}).

\section*{Acknowledgments}
A.P., L.T. and S.B. are partially supported by the PRIN-INAF 2011 with the 
project "Transient Universe: from ESO Large to PESSTO".
A.G. acknowledges support by the EU/FP7 via ERC grant n 307260, 
a GIF grant, and the Kimmel award.
The research of JRM is funded through a Royal Society University Research Fellowship.
Visiting Astronomer at the Infrared Telescope Facility, which is operated 
by the University of Hawaii under Cooperative Agreement no. NNX-08AE38A with 
the National Aeronautics and Space Administration, Science Mission Directorate, 
Planetary Astronomy Program. 
This paper is based on observations made with the following facilities:
ESO NTT Telescope ( program ID 188.D-3003 ), LCOGT 1m network of 
Telescopes (Haleaka, Siding Spring, Cerro Tololo, Mc Donald Observatory),
NASA Infrared Telescope Facility (IRTF), 1.8m and 1.2m Telescopes (Asiago)

\appendix
\begin{table*}
 \centering
  \begin{minipage}{140mm}
  \caption{Journal of spectroscopic observations.}
  \label{tab1}  
  \begin{tabular}{@{}ccccccc@{}}
  \hline   
Date  &   Telescope &  JD  & Phase\footnote{Relative to the date of the estimate 
shock breakout (JD =  2456497.0).} &
Instrument & Range & Resolution FWHM\footnote{FWHM of night-sky emission lines.} 
 \\
 &  & $-$2,400,000 (days) & (days) &  & (\AA) & (\AA) \\
 \hline
   2013 Jul  27  & IRTF     &  56501.1  &  4.1     & SpeX & 9000-25000 &    16/22/29 \\
  2013 Jul  27 &  FTS       &  56501.2   &  4.2     & FLOYDS & 3200-10000 &   5/12 \\
  2013 Jul   31 &  1.22m &  56504.6   &  7.6      & B$\&$C & 3400-8000   &   6.5 \\
  2013 Aug  01 &  FTN    &  56506     &  9.0     & FLOYDS & 3200-10000 &   6/13 \\
  2013 Aug  01 &  FTS    &  56506.2   &  9.2     & FLOYDS & 3200-10000 &   9/16 \\
  2013 Aug  02 &  FTN    &  56507     &  10      & FLOYDS & 3200-10000 &   6/13 \\
  2013 Aug  02 &  FTS    &  56507.3   &   10.3   & FLOYDS & 3200-10000 &   10/16 \\
  2013 Aug  03 &  FTN    &  56508     &  11      & FLOYDS & 3200-10000 &   6/13 \\
  2013 Aug  03 &  FTS    &  56508.2   &   11.2   & FLOYDS & 3200-10000 &   10/16 \\
  2013 Aug  04 &  FTS    &  56509.2   &   12.2   & FLOYDS & 3200-10000 &   9/15 \\
  2013 Aug  05 &  FTN    &  56510     &  12.9    & FLOYDS & 3200-10000 &   6/13 \\
  2013 Aug  05 &  FTS    &  56510.2   &  13.2    & FLOYDS & 3200-10000 &   10/17 \\
  2013 Aug  07 &  1.22m & 56511.6   &  14.6    & B$\&$C & 3400-8000   &   6.5 \\
  2013 Aug 08 &  1.22m & 56512.5   &   15.5    & B$\&$C & 3400-8000   &   6.5\\
  2013 Aug  08 &  FTS    &  56513.3   &  16.3    & FLOYDS & 3200-10000 &   10/16 \\
  2013 Aug  09 &  FTS    &  56514.1   &   17.1   & FLOYDS & 3200-10000 &   9/14 \\
  2013 Aug  10 &  FTS    &  56514.1   &   17.1   & FLOYDS & 3200-10000 &   9/14 \\
  2013 Aug  11 &  FTS    & 56516.2    &  19.2    & FLOYDS & 3200-10000 &   10/17  \\
  2013 Aug  12 &  FTS    & 56517.2  &   20.2    & FLOYDS & 3200-10000 &    11/16   \\
  2013 Aug  13 &  FTS    & 56518.1  &   21.1     & FLOYDS & 3200-10000 &   11/17   \\
  2013 Aug  15 &  FTS    &  56520.1 &  23.1    & FLOYDS & 3200-10000 &     10/16   \\
  2013 Aug  17 &  NTT   &  56521.7   &  24.7    & SOFI & 9000-25000 &   20/40 \\
  2013 Aug  17 &  FTS    &  56522.1   &  25.1    & FLOYDS & 3200-10000 &   10/16 \\
  2013 Aug  21 &  1.8m &  56526.42  &  29.4  &  Echelle &  5800-6000   &  0.4 \\
\hline
\end{tabular}
\end{minipage}
\end{table*}

\begin{table*}
\centering
 \begin{minipage}{140mm}
\caption{Photometric Data}
\begin{tabular}{ccccc|ccccc}
\hline
Date & JD & mag &  Filter & telescope$^a$ &Date & JD & mag &  Filter & telescope\footnote{1m0-08 (McDonald observatory, USA); 1m0-10, 1m0-12, 1m0-13 (Sutherland, South Africa), 1m0-04, 1m0-05, 1m0-09  (Cerro Tololo, Cile); 1m0-03, 1m0-11 (Siding Spring, Australia)} \\
\hline
2013-07-27 &56500.333 &13.077 0.014 &B & 1m0-04 & 2013-08-13 & 56517.391 & 12.896 0.018 & U & 1m0-04  \\
2013-07-27 &56500.338 &13.019 0.015 &V & 1m0-04 & 2013-08-13 & 56517.396 & 13.160 0.015 & B & 1m0-04  \\
2013-07-27 &56500.354 &12.946 0.018 &R & 1m0-04 & 2013-08-13 & 56517.398 & 12.550 0.015 & V & 1m0-04  \\
2013-07-27 &56500.357 &12.941 0.021 &I & 1m0-04 & 2013-08-13 & 56517.400 & 12.335 0.019 & R & 1m0-04  \\
2013-07-27 &56500.374 &12.198 0.016 &U & 1m0-04 & 2013-08-13 & 56517.402 & 12.197 0.023 & I & 1m0-04  \\
2013-07-28 &56501.333 &12.986 0.017 &B & 1m0-04 & 2013-08-14 & 56518.290 & 12.953 0.018 & U & 1m0-04  \\
2013-07-28 &56501.338 &12.914 0.019 &V & 1m0-04 & 2013-08-14 & 56518.295 & 13.196 0.015 & B & 1m0-04  \\
2013-07-28 &56501.354 &12.784 0.023 &R & 1m0-04 & 2013-08-14 & 56518.297 & 12.577 0.021 & V & 1m0-04  \\
2013-07-28 &56501.357 &12.764 0.026 &I & 1m0-04 & 2013-08-14 & 56518.299 & 12.404 0.027 & R & 1m0-04  \\
2013-07-28 &56501.374 &12.114 0.017 &U & 1m0-04 & 2013-08-14 & 56518.301 & 12.209 0.027 & I & 1m0-04  \\
2013-07-28 &56501.375 &13.015 0.017 &B & 1m0-08 & 2013-08-15 & 56519.283 & 12.989 0.022 & U & 1m0-09  \\
2013-07-29 &56502.333 &12.874 0.014 &B & 1m0-04 & 2013-08-15 & 56519.288 & 13.133 0.024 & B & 1m0-09  \\
2013-07-29 &56502.338 &12.798 0.016 &V & 1m0-09 & 2013-08-15 & 56519.290 & 12.643 0.032 & V & 1m0-09  \\
2013-07-29 &56502.354 &12.650 0.016 &R & 1m0-09 & 2013-08-15 & 56519.293 & 12.212 0.046 & I & 1m0-09  \\
2013-07-29 &56502.357 &12.621 0.019 &I & 1m0-09 & 2013-08-16 & 56520.286 & 13.214 0.016 & B & 1m0-09  \\
2013-08-02 &56506.358 &12.775 0.013 &B & 1m0-04 & 2013-08-16 & 56520.288 & 12.570 0.021 & V & 1m0-09  \\
2013-08-02 &56506.362 &12.593 0.014 &V & 1m0-04 & 2013-08-16 & 56520.290 & 12.198 0.037 & R & 1m0-09  \\
2013-08-03 &56507.356 &11.999 0.017 &U & 1m0-08 & 2013-08-16 & 56520.292 & 12.257 0.023 & I & 1m0-09  \\
2013-08-03 &56507.361 &12.399 0.022 &R & 1m0-08 & 2013-08-17 & 56521.278 & 13.169 0.020 & U & 1m0-05  \\
2013-08-03 &56507.363 &12.320 0.020 &I & 1m0-08 & 2013-08-17 & 56521.282 & 13.355 0.018 & B & 1m0-05  \\
2013-08-03 &56507.401 &12.703 0.021 &B & 1m0-08 & 2013-08-17 & 56521.285 & 12.642 0.017 & V & 1m0-05  \\
2013-08-03 &56507.404 &12.523 0.022 &V & 1m0-08 & 2013-08-17 & 56521.287 & 12.293 0.022 & R & 1m0-05  \\
2013-08-04 &56508.361 &12.009 0.017 &U & 1m0-08 & 2013-08-17 & 56521.288 & 12.268 0.026 & I & 1m0-05  \\
2013-08-04 &56508.366 &12.365 0.019 &R & 1m0-08 & 2013-08-18 & 56522.433 & 13.340 0.025 & U & 1m0-08  \\
2013-08-04 &56508.367 &12.298 0.020 &I & 1m0-08 & 2013-08-18 & 56522.438 & 13.395 0.020 & B & 1m0-08  \\
2013-08-04 &56508.371 &12.739 0.014 &B & 1m0-08 & 2013-08-18 & 56522.441 & 12.630 0.023 & V & 1m0-08  \\
2013-08-04 &56508.374 &12.535 0.018 &V & 1m0-08 & 2013-08-18 & 56522.442 & 12.276 0.026 & R & 1m0-08  \\
2013-08-05 &56509.353 &12.779 0.014 &B & 1m0-08 & 2013-08-18 & 56522.444 & 12.222 0.026 & I & 1m0-08  \\
2013-08-05 &56509.356 &12.558 0.018 &V & 1m0-08 & 2013-08-20 & 56524.340 & 13.538 0.023 & U & 1m0-05  \\
2013-08-05 &56509.429 &12.141 0.017 &U & 1m0-08 & 2013-08-20 & 56524.344 & 13.555 0.018 & B & 1m0-05  \\
2013-08-05 &56509.434 &12.344 0.023 &R & 1m0-08 & 2013-08-20 & 56524.347 & 12.745 0.022 & V & 1m0-05  \\
2013-08-05 &56509.436 &12.303 0.020 &I & 1m0-08 & 2013-08-20 & 56524.349 & 12.512 0.020 & R & 1m0-05  \\
2013-08-06 &56510.349 &12.782 0.020 &B & 1m0-08 & 2013-08-20 & 56524.351 & 12.344 0.020 & I & 1m0-05  \\
2013-08-06 &56510.352 &12.545 0.023 &V & 1m0-08 & 2013-08-21 & 56525.300 & 13.667 0.034 & U & 1m0-08  \\
2013-08-06 &56510.364 &12.215 0.022 &U & 1m0-08 & 2013-08-21 & 56525.305 & 13.590 0.022 & B & 1m0-08  \\
2013-08-06 &56510.369 &12.333 0.027 &R & 1m0-08 & 2013-08-21 & 56525.308 & 12.748 0.023 & V & 1m0-08  \\
2013-08-06 &56510.370 &12.268 0.025 &I & 1m0-08 & 2013-08-21 & 56525.309 & 12.534 0.028 & R & 1m0-08  \\
2013-08-09 &56513.337 &12.424 0.020 &U & 1m0-08 & 2013-08-21 & 56525.311 & 12.223 0.032 & I & 1m0-08  \\
2013-08-09 &56513.341 &12.316 0.023 &R & 1m0-08 & 2013-08-21 & 56525.314 & 13.658 0.021 & U & 1m0-09  \\
2013-08-09 &56513.343 &12.214 0.027 &I & 1m0-08 & 2013-08-21 & 56525.319 & 13.580 0.015 & B & 1m0-09  \\
2013-08-09 &56513.367 &12.920 0.015 &B & 1m0-08 & 2013-08-21 & 56525.321 & 12.739 0.017 & V & 1m0-09  \\
2013-08-09 &56513.370 &12.537 0.021 &V & 1m0-08 & 2013-08-21 & 56525.323 & 12.453 0.023 & R & 1m0-09  \\
2013-08-09 &56513.373 &12.489 0.016 &U & 1m0-04 & 2013-08-21 & 56525.324 & 12.231 0.018 & I & 1m0-09  \\
2013-08-09 &56513.378 &12.347 0.018 &R & 1m0-04 & 2013-08-22 & 56526.318 & 13.812 0.023 & U & 1m0-04  \\
2013-08-09 &56513.380 &12.252 0.023 &I & 1m0-04 &    2013-08-22  & 56526.321  & 13.666 0.015  & B   & 1m0-04   \\
2013-08-09 &56513.397 &12.932 0.015 &B & 1m0-04 &    2013-08-22  & 56526.324  & 12.788 0.020  & V   & 1m0-04   \\
2013-08-09 &56513.400 &12.594 0.018 &V & 1m0-04 &    2013-08-22  & 56526.326  & 12.516 0.023  & R   & 1m0-04   \\
2013-08-10 &56514.171 &12.914 0.016 &B & 1m0-10 &    2013-08-22  & 56526.327  & 12.238 0.021  & I   & 1m0-04	\\
2013-08-10 &56514.173 &12.579 0.015 &V & 1m0-10 &    2013-08-25  & 56529.295  & 14.087 0.027  & U   & 1m0-04	\\
2013-08-10 &56514.303 &12.493 0.072 &U & 1m0-09 &    2013-08-25  & 56529.300  & 13.829 0.016  & B   & 1m0-04	\\
2013-08-10 & 56514.307 & 12.240 0.020 & R & 1m0-09 & 2013-08-25  & 56529.303  & 12.920 0.016  & V   & 1m0-04	\\
2013-08-10 & 56514.309 & 12.293 0.023 & I & 1m0-09 & 2013-08-25  & 56529.305  & 12.579 0.019  & R   & 1m0-04	\\
2013-08-11 & 56515.171 & 13.011 0.015 & B & 1m0-13 & 2013-08-25  & 56529.306  & 12.465 0.026  & I   & 1m0-04 	\\
2013-08-11 & 56515.174 & 12.538 0.016 & V & 1m0-13 & 2013-08-29  & 56533.428  & 14.389 0.029  & U   & 1m0-08 	\\
2013-08-12 & 56516.291 & 13.089 0.014 & B & 1m0-04 & 2013-08-29  & 56533.433  & 14.005 0.022  & B   & 1m0-08  \\
2013-08-12 & 56516.294 & 12.566 0.014 & V & 1m0-04 & 2013-08-29  & 56533.435  & 12.927 0.020  & V   & 1m0-08 \\
 \hline
\end{tabular}
\label{tab2}
\end{minipage}
\end{table*}

\begin{table*}
\centering
\small
 \begin{minipage}{140mm}
\caption{...continued}
\begin{tabular}{ccccc|ccccc}
\hline
Date & JD & mag &  Filter & telescope$^a$ &Date & JD & mag &  Filter & telescope\footnote{1m0-08 
(McDonald observatory, USA); 1m0-10, 1m0-12, 1m0-13 (Sutherland, South Africa), 1m0-04, 1m0-05, 
1m0-09  (Cerro Tololo, Cile); 1m0-03, 1m0-11 (Siding Spring, Australia)} \\
\hline
2013-08-29  & 56533.437  & 12.608 0.028  & R   & 1m0-08  &    2013-08-17  & 56521.288  & 12.652 0.031  & zs  & 1m0-04     \\
2013-08-29  & 56533.439  & 12.416 0.031  & I   & 1m0-08  &    2013-08-18  & 56522.357  & 13.035 0.020  & gp  & 1m0-08     \\  
2013-08-30  & 56534.323  & 14.428 0.027  & U   & 1m0-08  &   2013-08-18  & 56522.359  & 12.513 0.019  & rp  & 1m0-08      \\
2013-08-30  & 56534.328  & 14.028 0.019  & B   & 1m0-08  &   2013-08-18  & 56522.361  & 12.590 0.020  & ip  & 1m0-08      \\
2013-08-30  & 56534.331  & 12.957 0.022  & V   & 1m0-08 &    2013-08-18  & 56522.362  & 12.617 0.024  & zs  & 1m0-08     \\
2013-07-27  & 56500.341  & 12.980 0.015  & gp  & 1m0-04 &    2013-08-20  & 56524.327  & 14.841 0.035  & up  & 1m0-05    \\
2013-07-27  & 56500.346  & 13.079 0.019  & rp  & 1m0-04 &    2013-08-20  & 56524.331  & 13.175 0.017  & gp  & 1m0-05   \\
2013-07-27  & 56500.349  & 13.272 0.019  & ip  & 1m0-04 &    2013-08-20  & 56524.334  & 12.615 0.018  & rp  & 1m0-05   \\
2013-07-27  & 56500.360  & 13.418 0.024  & zs  & 1m0-04	 &   2013-08-20  & 56524.336  & 12.631 0.022  & ip  & 1m0-05    \\
2013-07-28  & 56501.346  & 12.886 0.022  & rp  & 1m0-04	 &   2013-08-21  & 56525.307  & 14.782 0.046  & up  & 1m0-04    \\
2013-07-28  & 56501.349  & 13.071 0.022  & ip  & 1m0-04	 &   2013-08-21  & 56525.311  & 13.191 0.017  & gp  & 1m0-04   	   \\
2013-07-28  & 56501.360  & 13.228 0.033  & zs  & 1m0-04	 &   2013-08-21  & 56525.314  & 12.568 0.019  & rp  & 1m0-04     \\
2013-07-29  & 56502.341  & 12.774 0.014  & gp  & 1m0-09	 &   2013-08-21  & 56525.316  & 12.613 0.020  & ip  & 1m0-04    \\
2013-07-29  & 56502.346  & 12.835 0.017  & rp  & 1m0-09	 &   2013-08-21  & 56525.317  & 12.636 0.024  & zs  & 1m0-04   	   \\
2013-07-29  & 56502.349  & 12.927 0.021  & ip  & 1m0-09	 &   2013-08-22  & 56526.304  & 15.066 0.072  & up  & 1m0-09   	   \\
2013-07-29  & 56502.360  & 13.038 0.023  & zs  & 1m0-09	 &   2013-08-22  & 56526.311  & 12.594 0.027  & rp  & 1m0-09   	   \\
2013-08-02  & 56506.363  & 12.693 0.013  & gp  & 1m0-04	 &   2013-08-22  & 56526.313  & 12.618 0.023  & ip  & 1m0-09   	   \\
2013-08-02  & 56506.366  & 12.590 0.014  & rp  & 1m0-04	 &   2013-08-22  & 56526.314  & 12.635 0.033  & zs  & 1m0-09   	   \\
2013-08-02  & 56506.369  & 12.760 0.015  & ip  & 1m0-04	 &   2013-08-25  & 56529.296  & 15.080 0.030  & up  & 1m0-09   	   \\
2013-08-03  & 56507.405  & 12.702 0.057  & gp  & 1m0-08	 &   2013-08-25  & 56529.300  & 13.377 0.020  & gp  & 1m0-09   	   \\
2013-08-03  & 56507.408  & 12.481 0.020  & rp  & 1m0-08	 &   2013-08-25  & 56529.303  & 12.646 0.023  & rp  & 1m0-09   	   \\
2013-08-03  & 56507.410  & 12.640 0.019  & ip  & 1m0-08	 &   2013-08-25  & 56529.304  & 12.679 0.026  & ip  & 1m0-09   	   \\
2013-08-04  & 56508.369  & 12.741 0.024  & zs  & 1m0-08	 &   2013-08-25  & 56529.306  & 12.666 0.027  & zs  & 1m0-09   	   \\
2013-08-04  & 56508.378  & 12.492 0.018  & rp  & 1m0-08	 &   2013-08-29  & 56533.459  & 13.494 0.024  & gp    & 1m0-08 	   \\
2013-08-04  & 56508.380  & 12.690 0.022  & ip  & 1m0-08	 &   2013-08-29  & 56533.461  & 12.750 0.024  & rp    & 1m0-08 	   \\
2013-08-05  & 56509.357  & 12.593 0.041  & gp  & 1m0-08	 &   2013-08-29  & 56533.463  & 12.753 0.022  & ip    & 1m0-08 	   \\
2013-08-05  & 56509.360  & 12.473 0.022  & rp  & 1m0-08	 &   2013-08-29  & 56533.465  & 12.759 0.032  & zs    & 1m0-08 	   \\
2013-08-05  & 56509.437  & 12.717 0.027  & zs  & 1m0-08	 &   2013-07-31  & 56504.481  & 12.238 0.059  & uvm2  & swift  	   \\
2013-08-05  & 56509.448  & 12.591 0.021  & ip  & 1m0-08	 &   2013-07-31  & 56504.473  & 11.717 0.057  & uvw1  & swift  	   \\
2013-08-06  & 56510.353  & 12.592 0.041  & gp  & 1m0-08	 &   2013-07-31  & 56504.476  & 12.538 0.070  & uvw2  & swift  	   \\
2013-08-06  & 56510.356  & 12.463 0.023  & rp  & 1m0-08	 &   2013-08-01  & 56505.006  & 12.149 0.058  & uvm2  & swift  \\
2013-08-06  & 56510.358  & 12.575 0.023  & ip  & 1m0-08	 &   2013-08-01  & 56504.993  & 11.752 0.056  & uvw1  & swift  \\
2013-08-06  & 56510.372  & 12.640 0.026  & zs  & 1m0-08	 &   2013-08-01  & 56504.999  & 12.505 0.069  & uvw2  & swift  \\
2013-08-09  & 56513.345  & 12.667 0.026  & zs  & 1m0-08	 &   2013-08-01  & 56505.362  & 12.259 0.058  & uvm2  & swift  \\
2013-08-09  & 56513.371  & 12.647 0.024  & gp  & 1m0-08	 &   2013-08-01  & 56505.335  & 11.839 0.057  & uvw1  & swift  \\
2013-08-09  & 56513.374  & 12.395 0.022  & rp  & 1m0-08	 &   2013-08-01  & 56505.319  & 12.585 0.071  & uvw2  & swift  \\
2013-08-09  & 56513.376  & 12.522 0.021  & ip  & 1m0-08	 &   2013-08-03  & 56507.499  & 12.734 0.075  & uvm2  & swift  \\
2013-08-09  & 56513.381  & 12.676 0.027  & zs  & 1m0-04	 &   2013-08-03  & 56507.582  & 13.098 0.070  & uvw2  & swift  \\
2013-08-09  & 56513.401  & 12.706 0.019  & gp  & 1m0-04	 &   2013-08-05  & 56509.381  & 13.413 0.070  & uvw2  & swift  \\
2013-08-09  & 56513.404  & 12.500 0.018  & rp  & 1m0-04	 &   2013-08-07  & 56511.734  & 13.000 0.059  & uvw1  & swift  \\
2013-08-09  & 56513.406  & 12.572 0.027  & ip  & 1m0-04	 &   2013-08-07  & 56511.953  & 13.962 0.072  & uvw2  & swift  \\
2013-08-10  & 56514.178  & 12.433 0.019  & rp  & 1m0-10	 &   2013-08-08  & 56512.477  & 13.235 0.087  & uvw1  & swift  \\
2013-08-10  & 56514.179  & 12.514 0.022  & ip  & 1m0-10	 &   2013-08-08  & 56512.490  & 14.005 0.060  & uvm2  & swift  \\
2013-08-10  & 56514.310  & 12.594 0.024  & zs  & 1m0-09	 &   2013-08-08  & 56512.482  & 13.979 0.113  & uvw2  & swift  \\
2013-08-11  & 56515.175  & 12.743 0.015  & gp  & 1m0-13	 &   2013-08-08  & 56512.687  & 14.154 0.070  & uvw2  & swift  \\
2013-08-11  & 56515.178  & 12.482 0.016  & rp  & 1m0-13	 &   2013-08-09  & 56513.669  & 14.490 0.071  & uvw2  & swift  \\
2013-08-11  & 56515.179  & 12.575 0.018  & ip  & 1m0-13	 &   2013-08-09  & 56513.688  & 14.411 0.071  & uvw2  & swift  \\
2013-08-12  & 56516.296  & 12.791 0.014  & gp  & 1m0-04	 &   2013-08-09  & 56513.801  & 13.387 0.089  & uvw1  & swift  \\
2013-08-12  & 56516.298  & 12.470 0.014  & rp  & 1m0-04	 &   2013-08-09  & 56513.806  & 14.464 0.129  & uvw2  & swift  \\
2013-08-12  & 56516.300  & 12.562 0.018  & ip  & 1m0-04	 &   2013-08-12  & 56516.277  & 14.070 0.107  & uvw1  & swift  \\
2013-08-13  & 56517.312  & 13.820 0.049  & up  & 1m0-04	 &   2013-08-12  & 56516.282  & 15.261 0.168  & uvm2  & swift  \\
2013-08-13  & 56517.320  & 12.458 0.014  & rp  & 1m0-04	 &   2013-08-12  & 56516.290  & 15.197 0.065  & uvm2  & swift  \\
2013-08-13  & 56517.322  & 12.539 0.015  & ip  & 1m0-04	 &   2013-07-31  & 56504.474  & 11.968 0.241  & U     & swift  \\
2013-08-13  & 56517.323  & 12.645 0.019  & zs  & 1m0-04	 &   2013-07-31  & 56504.475  & 12.714 0.062  & B     & swift  \\
2013-08-14  & 56518.307  & 12.815 0.021  & gp  & 1m0-04	 &   2013-07-31  & 56504.478  & 12.694 0.043  & V     & swift  \\
2013-08-15  & 56519.294  & 12.814 0.023  & gp  & 1m0-04	 &   2013-08-01  & 56504.995  & 11.991 0.282  & U     & swift  \\
2013-08-15  & 56519.297  & 12.478 0.023  & rp  & 1m0-04	 &   2013-08-01  & 56504.996  & 12.700 0.045  & B     & swift  \\
2013-08-15  & 56519.308  & 12.836 0.018  & gp  & 1m0-08	 &   2013-08-02  & 56505.002  & 12.634 0.037  & V     & swift  \\
2013-08-16  & 56520.287  & 12.926 0.014  & gp  & 1m0-05	 &   2013-08-08  & 56512.481  & 12.747 0.096  & B     & swift  \\
2013-08-16  & 56520.289  & 12.499 0.018  & rp  & 1m0-05	 &   2013-08-08  & 56512.484  & 12.449 0.065  & V     & swift  \\
2013-08-16  & 56520.291  & 12.565 0.021  & ip  & 1m0-05	 &   2013-08-08  & 56512.668  & 12.210 0.049  & U     & swift  \\
2013-08-16  & 56520.292  & 12.693 0.024  & zs  & 1m0-05	 &     2013-08-09  & 56513.805  & 12.948 0.090  & B     & swift \\
2013-08-17  & 56521.282  & 12.954 0.021  & gp  & 1m0-04	 &     2013-08-09  & 56513.808  & 12.542 0.067  & V     & swift \\
 2013-08-17  & 56521.285  & 12.479 0.024  & rp  & 1m0-04  &  2013-08-12  & 56516.281  & 12.991 0.083  & B     & swift \\
 2013-08-17  & 56521.287  & 12.550 0.026  & ip  & 1m0-04  &  2013-08-12  & 56516.284  & 12.596 0.066  & V     & swift \\
 \hline
\end{tabular}
\label{tab2}
\end{minipage}
\end{table*}

\end{document}